\documentclass{PoS}

\usepackage{graphicx}
\usepackage{amsmath, amsthm, amssymb}

\newcommand{\mev}{\mathrm{MeV}}

\newcommand{\mevm}{\mathrm{MeV}/c^2}

\newcommand{\uu}{\mu^+\mu^-}
\newcommand{\pp}{\pi^+\pi^-}

\newcommand{\U}{\Upsilon}
\newcommand{\Uf}{\Upsilon(5S)}
\newcommand{\Uo}{\Upsilon(1S)}

\newcommand{\Ut}{\Upsilon(2S)}
\newcommand{\Uth}{\Upsilon(3S)}
\newcommand{\Un}{\Upsilon(nS)}

\newcommand{\hb}{h_b(1P)}
\newcommand{\hbp}{h_b(2P)}

\newcommand{\hbm}{h_b(mP)}
\newcommand{\jp}{J/\psi}

\newcommand{\pip}{\pi^{+}}
\newcommand{\pipm}{\pi^{\pm}}
\newcommand{\pimp}{\pi^{\mp}}

\newcommand{\br}{\mathcal{B}}

\newcommand{\zb}{Z_b}
\newcommand{\zbo}{Z_b(10610)}
\newcommand{\zbt}{Z_b(10650)}

\title{\boldmath Status and new results on the $\zb$ resonances}

\ShortTitle{Belle results in quarkonium}

\author{Alexander Bondar\\
  Budker Institute of Nuclear Physics SB RAS and Novosibirsk State University, Novosibirsk\\
  E-mail: \email{A.E.Bondar@inp.nsk.su}}
\author{\speaker{Roman Mizuk}\\
  Institute for Theoretical and Experimenal Physics, Moscow\\
  E-mail: \email{mizuk@itep.ru}}

\abstract{ Recently Belle observed two charged bottomonium-like
  states, $\zbo$ and $\zbt$, that are produced in the
  $\Uf\to\zb^{\pm}\pimp$ transitions and that decay to $\Un\pipm$
  ($n=1,2,3$) and $\hbm\pipm$ ($m=1,2$) channels. The masses of these
  states are close to the $B\bar{B}^*$ and $B^*\bar{B}^*$ thresholds,
  and their favored spin-parities are $J^P=1^+$. We review status and
  new results on the $\zb$ states, that include the observation of the
  $\zb\to B^{(*)}\bar{B}^*$ decays and an evidence for the neutral
  member of the $\zb$ isotriplet. All properties of the $\zb$ states
  are consistent with their molecular interpretation. }
 
\FullConference{Xth Quark Confinement and the Hadron Spectrum,\\
                October 8-12, 2012\\
                TUM Campus, Garching, Munich, Germany}

\begin{document}

Among unexpected results of B-factories is the observation of many
quarkonium-like states with properties that do not fit potential model
predictions. The first among them was $X(3872)$, observed by Belle in
2003~\cite{x3872_belle_1st}. Its mass is very close to the
$D^0\bar{D}^{*0}$ threshold and its favored interpretation is a
mixture of the conventional $\chi_{c1}(2P)$ charmonium with a
$D\bar{D}^*$ molecule. Other anomalous $XYZ$ states lie in the above
threshold region; they are discussed in a separate talk at this
meeting.

Recently Belle observed bottomonium-like states $\zbo$ and $\zbt$ with
masses close to the $B\bar{B}^*$ and $B^*\bar{B}^*$ thresholds,
respectively~\cite{zb_belle}. Unlike $X(3872)$, these states are
charged, i.e. their quark content is explicitly exotic, e.g.
$|b\bar{b}u\bar{d}\rangle$. We review the status and the most recent
results on $\zb$, including the $\zb$ decays to the $B\bar{B}^*$ and
$B^*\bar{B}^*$ channels and an evidence for their neutral isospin
partner $\zb^0$.

\section{\boldmath Observation of $\zb$ states in the $\Un\pp$ and $\hbm\pp$ channels}

Recently Belle observed the $\hb$ and $\hbp$ states in the
transitions $\Uf\to \hbm\pp$~\cite{hb_belle}. The rates of these
transitions appeared to be unsuppressed relative to the $\Uf\to\Un\pp$
($n=1,2,3$). The $\hbm$ production involves spin-flip of $b$-quark and
is suppressed as $(\Lambda_{QCD}/m_b)^2$ in the multipole expansion;
this unexpected result motivated further studies of the $\hbm$ and
$\Un$ production mechanisms.

\begin{figure}[tbhp]
\includegraphics[width=0.33\linewidth]{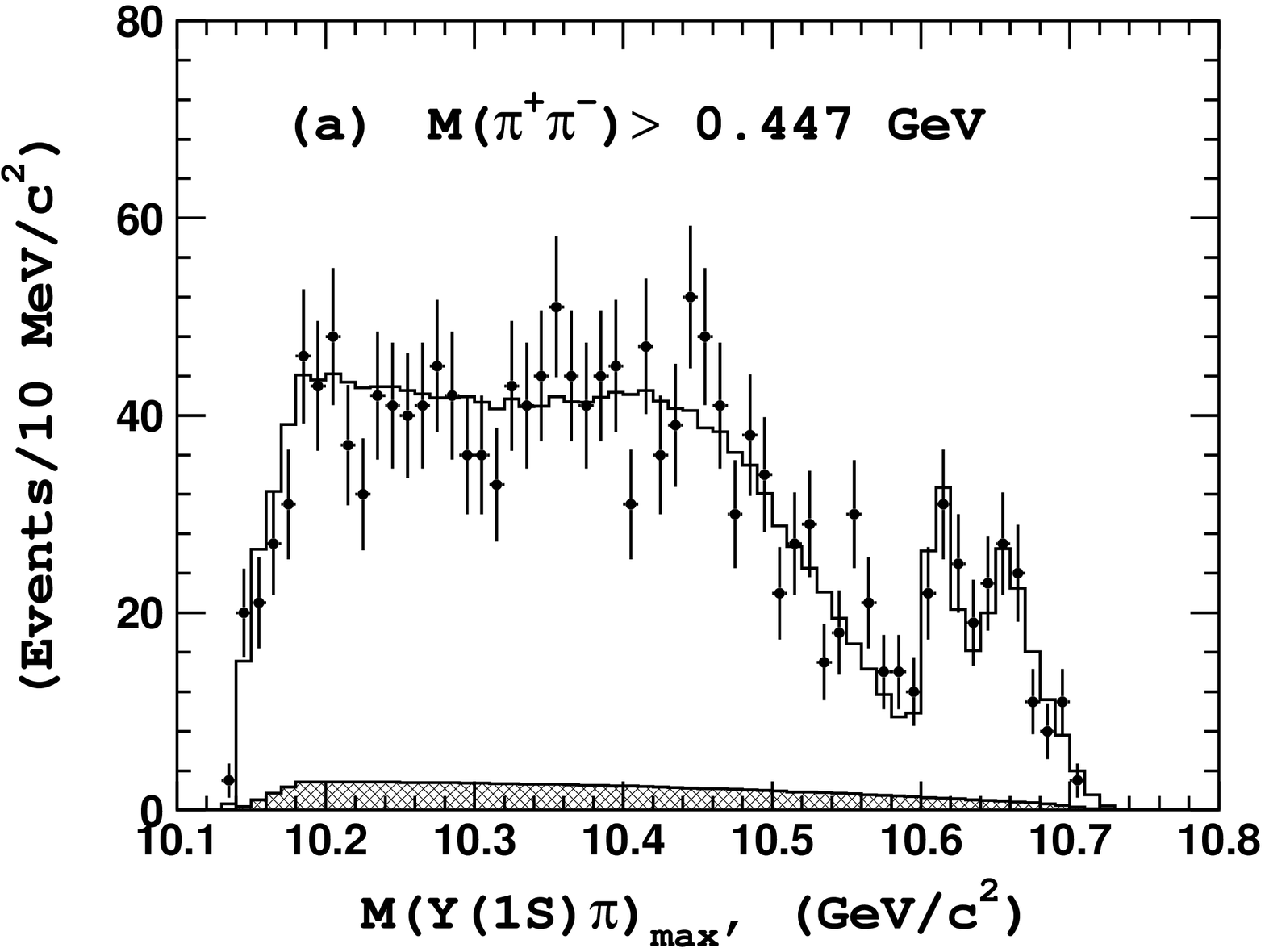}
\includegraphics[width=0.33\linewidth]{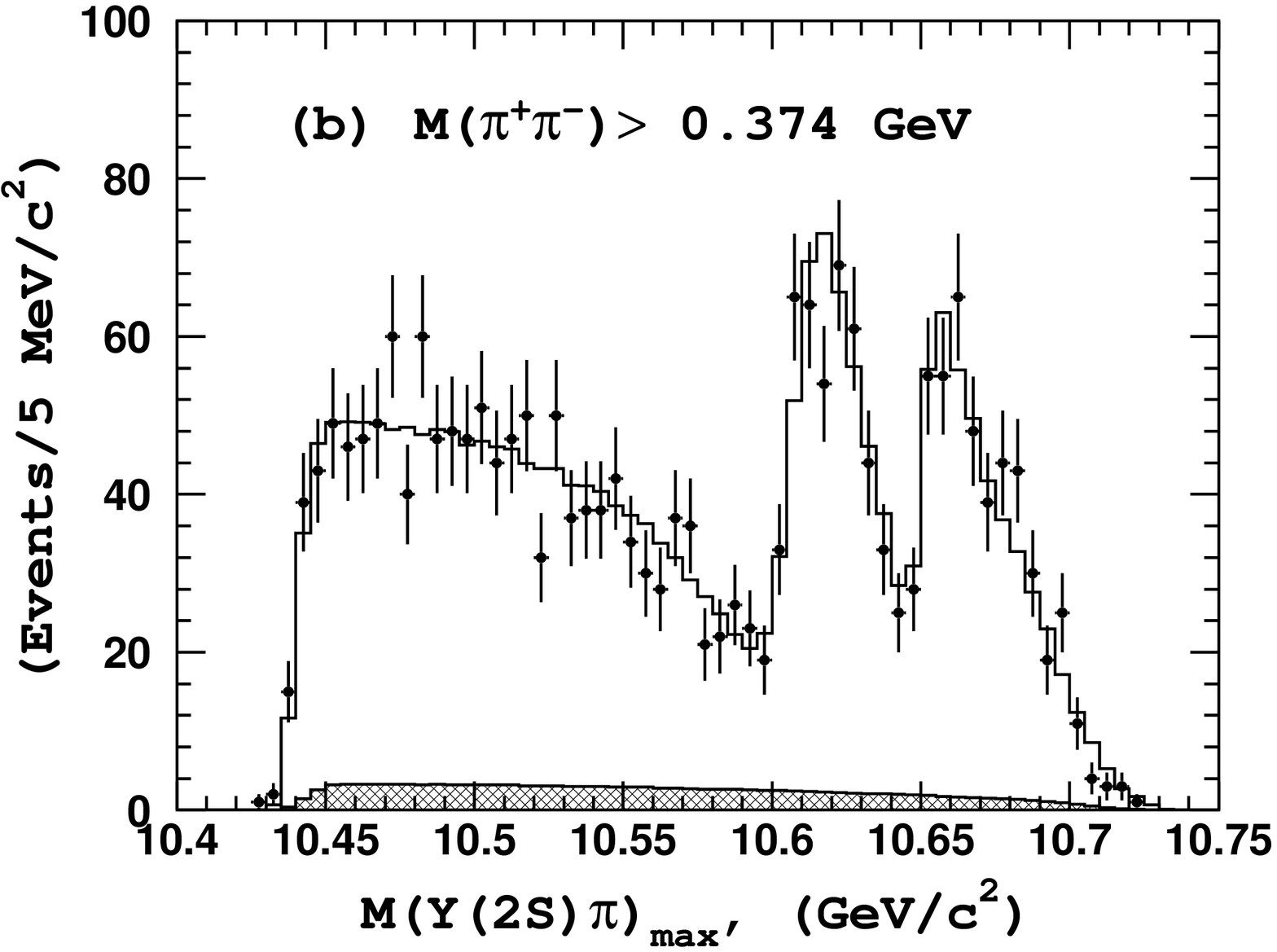}
\includegraphics[width=0.33\linewidth]{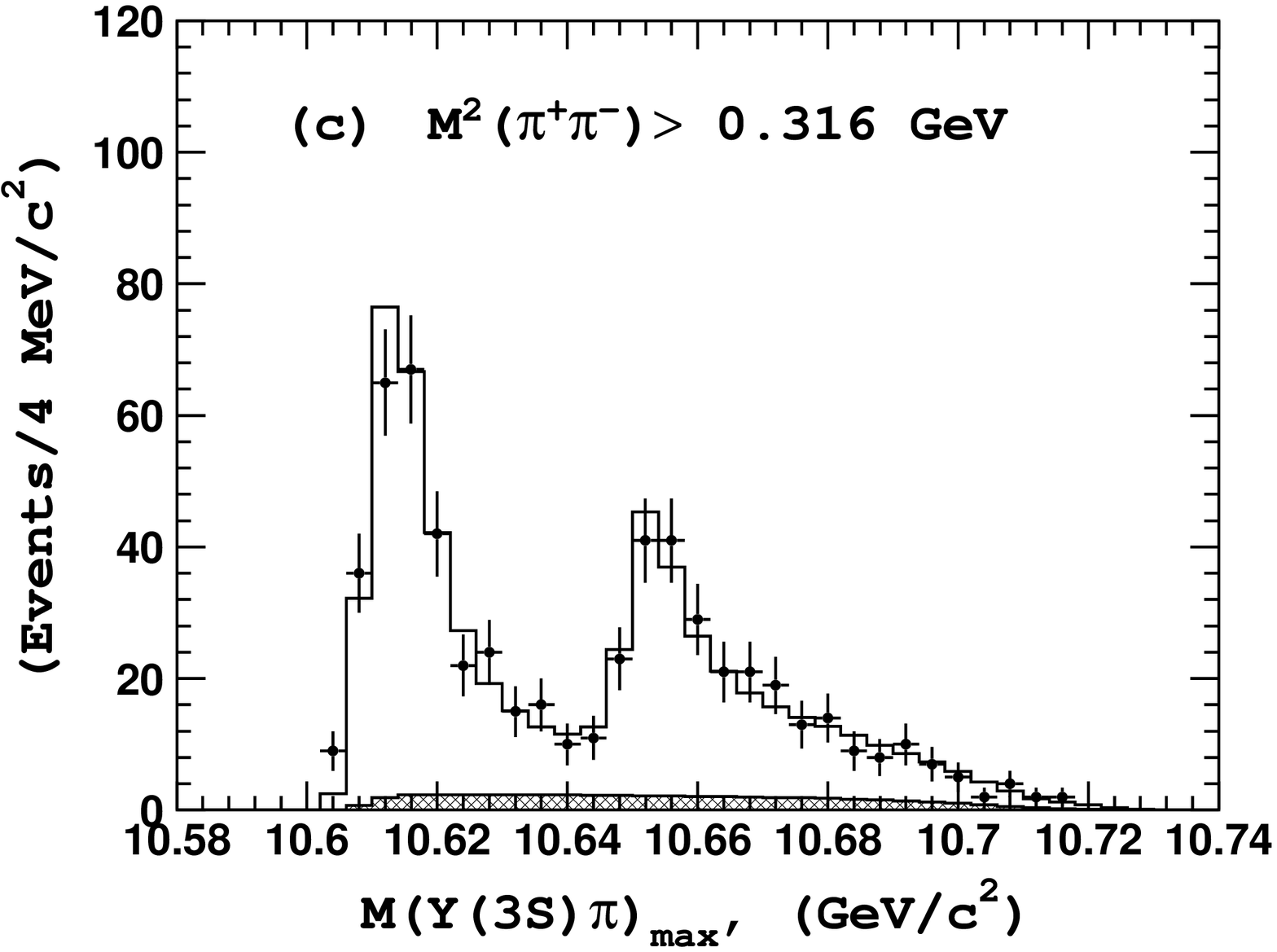}
\center
\includegraphics[width=0.33\linewidth]{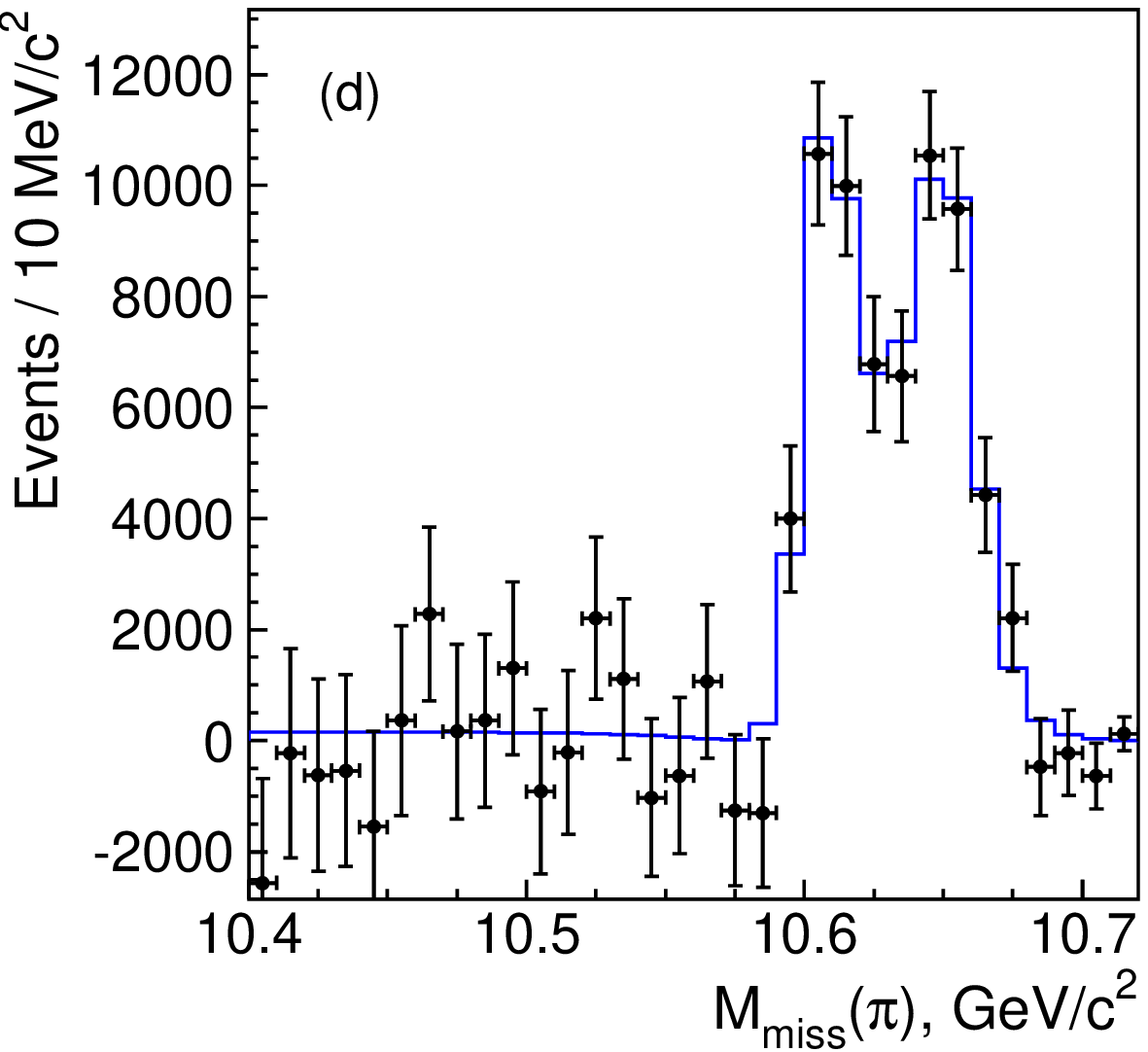}
\includegraphics[width=0.33\linewidth]{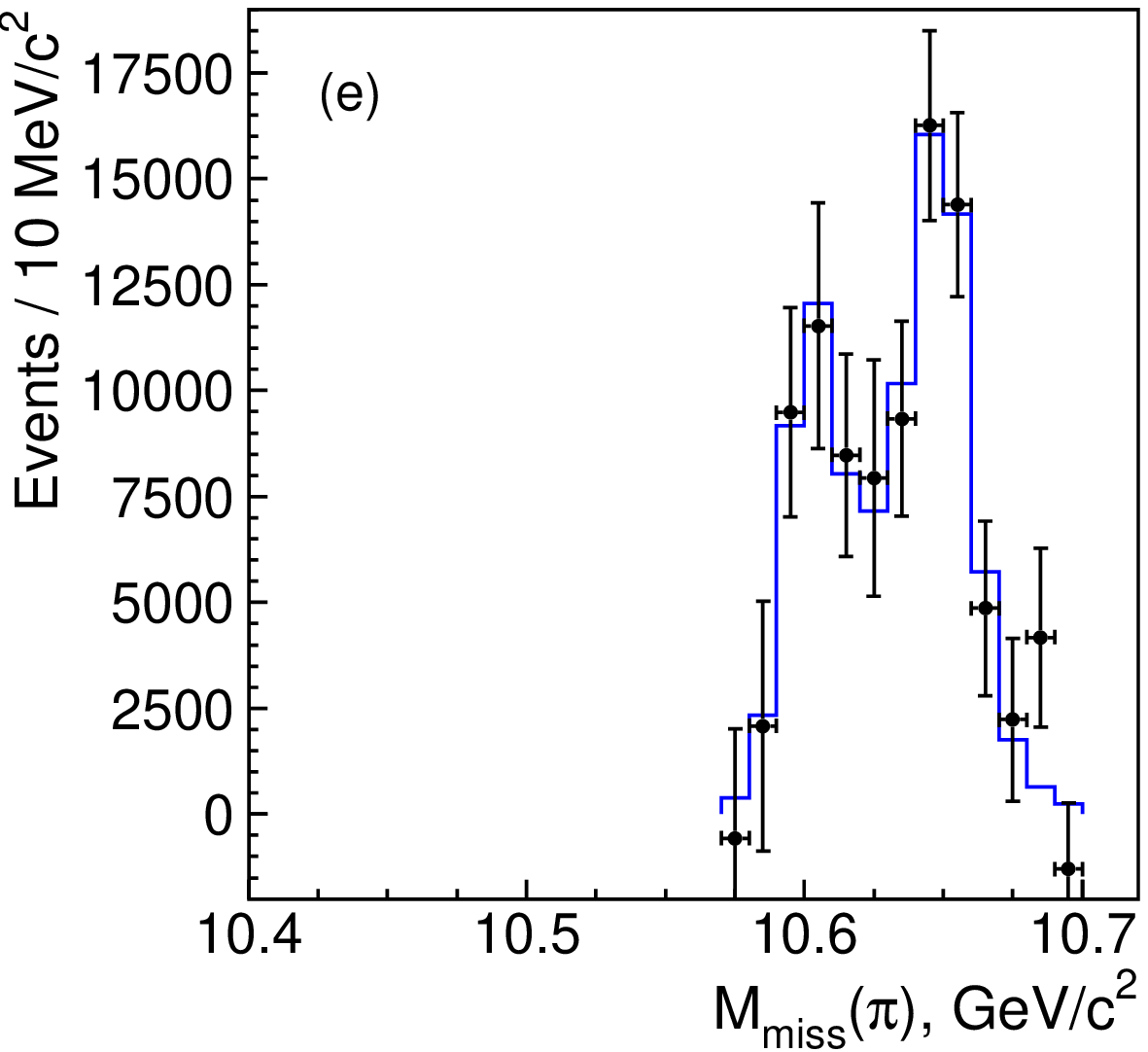}
\caption{ Invariant mass spectra of the (a) $\Uo\pipm$, (b)
  $\Ut\pipm$, (c) $\Uth\pipm$, (d) $\hb\pipm$ and (e) $\hbp\pipm$
  combinations. }
\label{fig:zb_signals}
\end{figure}

\begin{figure}[tbhp]
\center
\includegraphics[width=0.7\linewidth]{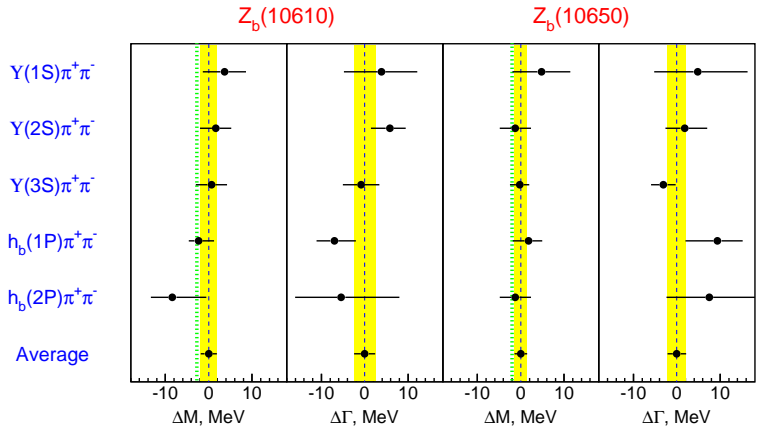}
\caption{ The deviations of the mass and width measurements of the
  $\zbo$ and $\zbt$ in different channels from the averaged over all
  channels value. Green vertical lines indicate the $B\bar{B}^*$ and
  $B^*\bar{B}^*$ thresholds.}
\label{fig:zb_table}
\end{figure}

Belle studied the resonant structure of the $\Uf\to\Un\pp$ and
$\hbm\pp$ decays ($n=1,2,3$; $m=1,2$)~\cite{zb_belle}. The $\Un$
[$\hbm$] states are reconstructed in the $\uu$ channel [inclusively
  using missing mass of the $\pp$ pairs].  Invariant mass spectra of
the $\Un\pipm$ and $\hbm\pipm$ combinations are shown in
Fig.~\ref{fig:zb_signals}.
Each distribution shows two peaks. For the channels $\Un\pp$
[$\hbm\pp$] the Dalitz plot analysis [fit to one-dimensional
  distributions] is performed.  The non-resonant contributions in the
$\hbm\pp$ channels are negligible, justifying the one-dimensional
analysis. Preliminary results of the angular analysis indicate that
both states have the same spin-parity
$J^P=1^+$~\cite{zb_belle_angular}, therefore coherent sum of
Breit-Wigner amplitudes is used to describe the signals. The Dalitz
plot model for the $\Uf\to\Un\pp$ channels includes also the $\pp$
resonances $f_0(980)$ and $f_2(1270)$, and non-resonant contribution,
parameterized as $a+b\,M_{\pp}^2$, where $a$ and $b$ are complex
numbers floating in the fit.
The masses and widths of the two peaks are found to be in good
agreement among different channels (see Fig.~\ref{fig:zb_table}).
Averaged over the five decay channels parameters are
\begin{align*}
M_1 & =(10607.4\pm2.0)\,\mevm, & M_2 & =(10652.2\pm1.5)\,\mevm, \\  
\Gamma_1 & =(18.4\pm2.4)\,\mev, & \Gamma_2 & =(11.5\pm2.2)\,\mev.
\end{align*}
The peaks are identified as signals of two new states, named $\zbo$
and $\zbt$. 

Another result of the amplitude analyses is that the phase between the
$\zbo$ and $\zbt$ amplitudes is zero for the $\Un\pp$ channels, and
$180^{\circ}$ for the $\hbm$ channels.

The masses of the $\zbo$ and $\zbt$ states are close to the
$B\bar{B}^*$ and $B^*\bar{B}^*$ thresholds, respectively. All the
properties of the $\zbo$ and $\zbt$ find natural explanation once
molecular structure for these states is assumed without even the need
of dynamic model. Considering the heavy-quark spin structure of the
$B^{(*)}\bar{B}^*$ molecule with $I^G(J^P)=1^+(1^+)$, one concludes
that $\zb$ contain both ortho- and para-bottomonium
components~\cite{zb_voloshin}. The weight of these components is
equal, therefore the decay to the $\hbm\pipm$ is not suppressed
relative to the $\Un\pipm$. The $\zbo$ and $\zbt$ differ by the sign
between ortho- and para-bottomonium components, this explains why the
$\zbo$ and $\zbt$ amplitudes appear with the sign plus for the
$\Un\pp$ channels and with the sign minus for the $\hbm\pp$
channels. In the limit of infinitely heavy $b$ quark the $B$ and $B^*$
mesons have equal mass, thus the $\zbo$ and $\zbt$ are also
degenerate. Given minus sign between the $\zb$ amplitudes in the
$\hbm\pp$ channel the contribution of this channel vanishes if the
heavy quark symmetry is exact.

\section{Observation of the $\zbo\to B\bar{B}^*$ and $\zbt\to B^*\bar{B}^*$ decays \label{sec:obs}}

\begin{figure}[tbhp]
\includegraphics[width=0.33\linewidth]{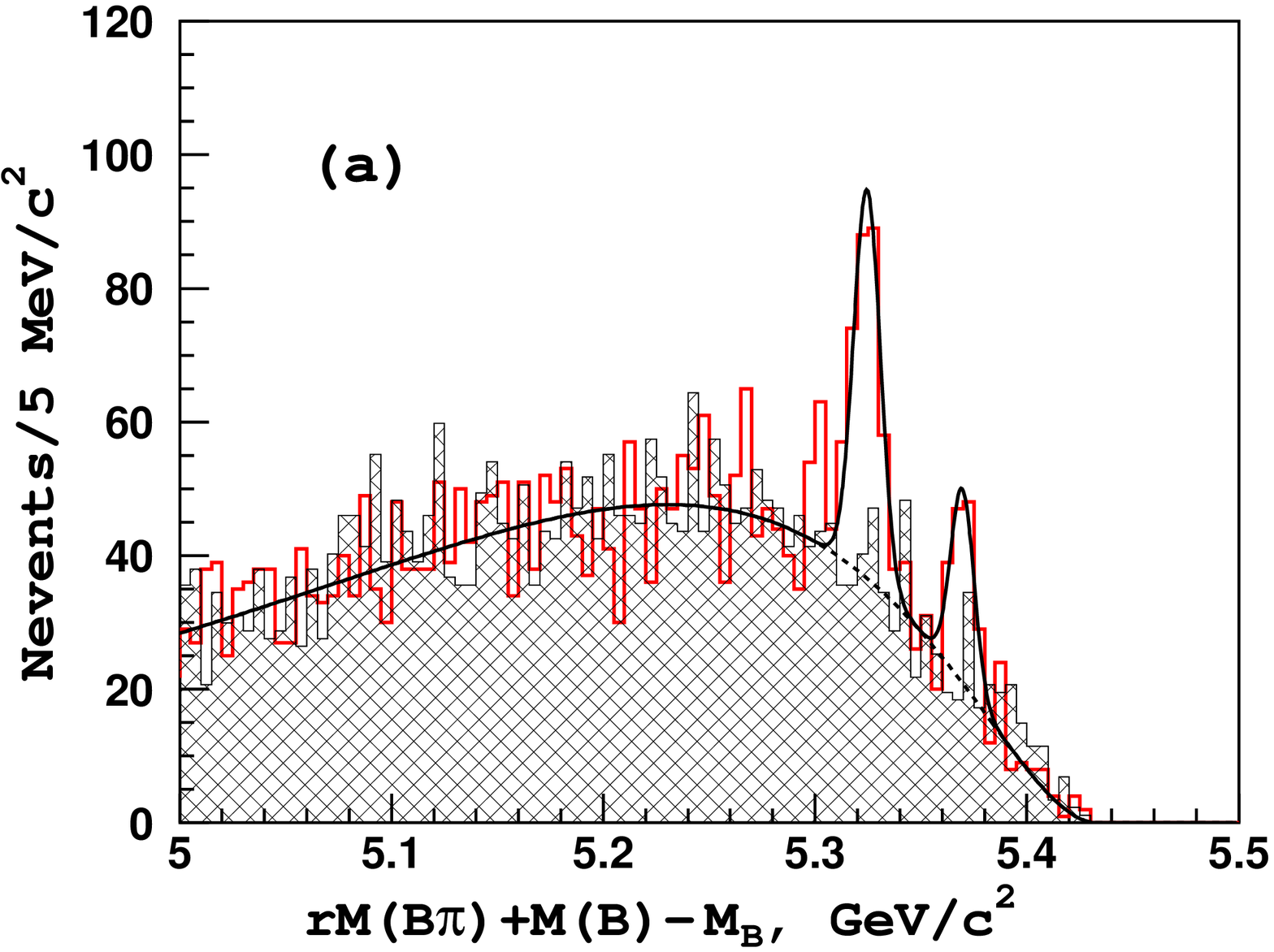}
\includegraphics[width=0.33\linewidth]{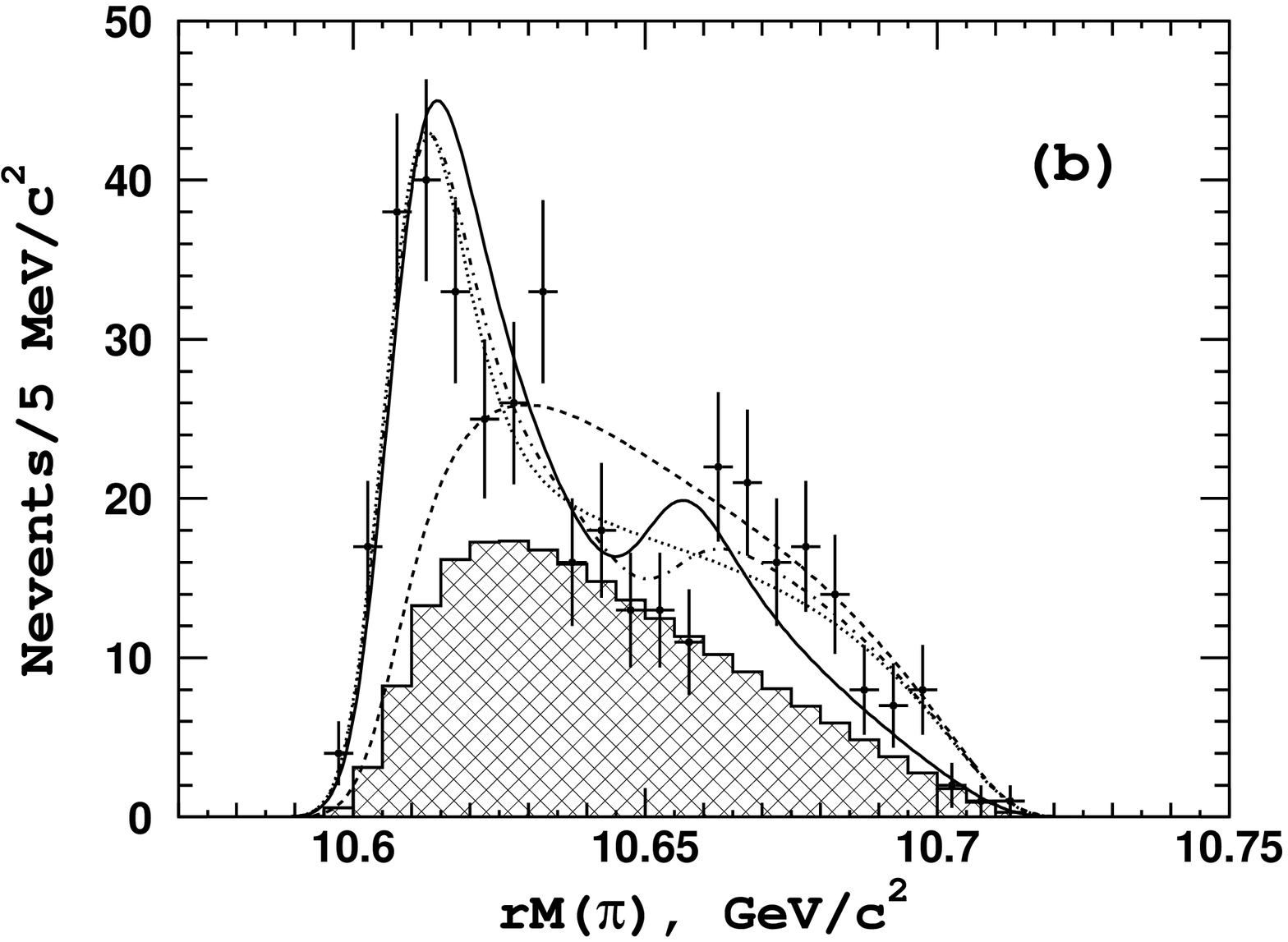}
\includegraphics[width=0.33\linewidth]{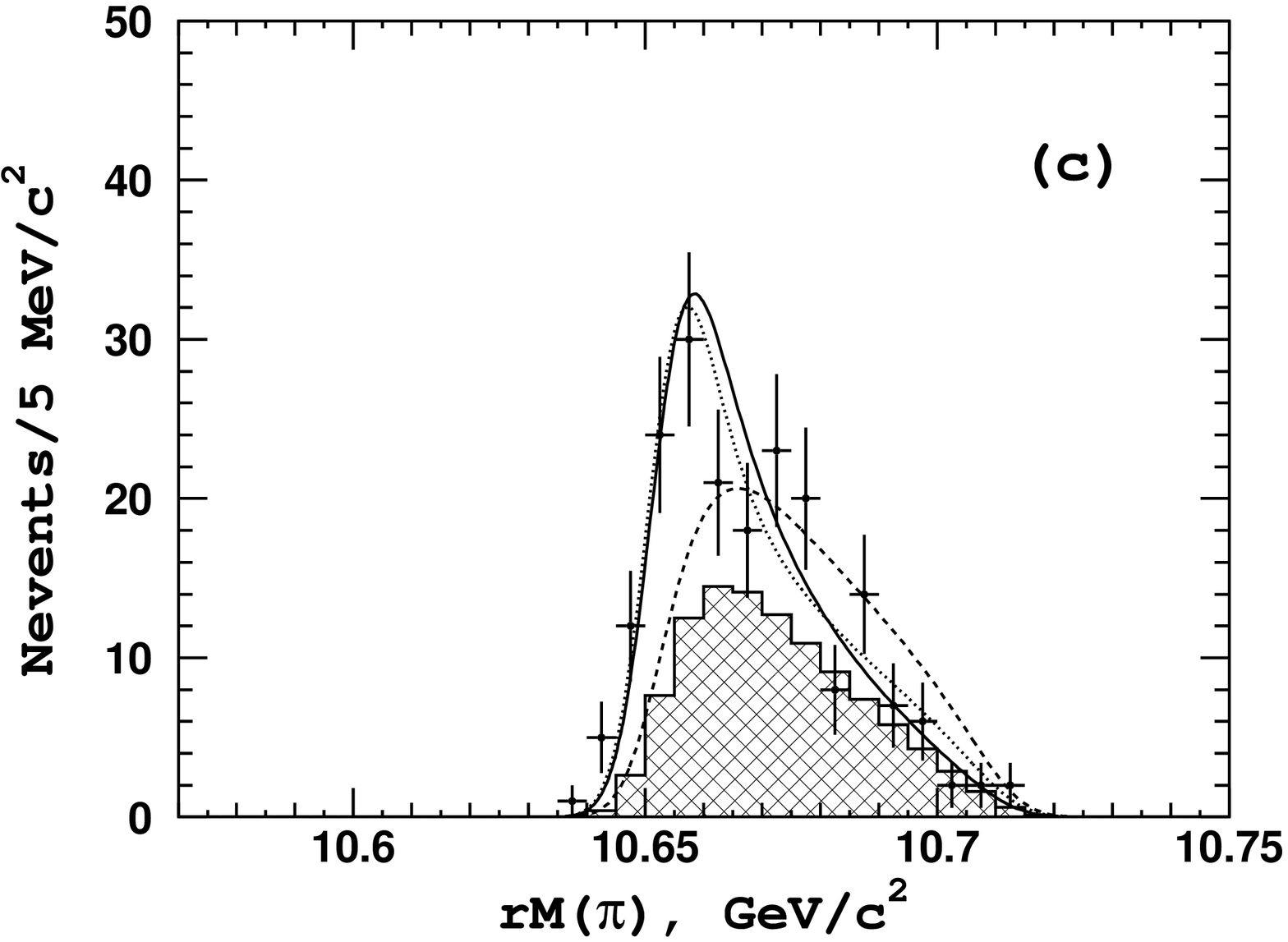}
\caption{ Missing mass of the pairs formed from the reconstructed $B$
  candidate and charged pion (a) and missing mass of the charged pions
  for the $B\pi$ combinations for (b) $\Uf\to B\bar{B}^*\pi$ and (c)
  $\Uf\to B^*\bar{B}^*\pi$ candidate events.}
\label{fig:bbpi}
\end{figure}

Given proximity to the thresholds and finite widths, it is natural to
expect that the rates of the ``fall-apart'' decays $\zbo\to
B\bar{B}^*$ and $\zbt\to B^*\bar{B}^*$ are substantial in the
molecular picture. To search for these transitions Belle studied the
$\Uf\to[B^{(*)}\bar{B}^*]^{\pm}\pi^{\mp}$
decays~\cite{zb_belle_bb}. One $B$ meson is reconstructed fully using
the $D^{(*)}\pi^+$ and $\jp K^{(*)}$ channels. The distribution of the
missing mass of the $B\pi^{\pm}$ pairs shows clear signals of the
$\Uf\to[B\bar{B}^*]^{\pm}\pi^{\mp}$ and
$\Uf\to[B^*\bar{B}^*]^{\pm}\pi^{\mp}$ decays [see Fig.~\ref{fig:bbpi}
  (a)];
corresponding branching fractions of $(2.83\pm0.29\pm0.46)\,\%$ and
$(1.41\pm0.19\pm0.24)\,\%$, respectively, are in agreement with
previous Belle measurement~\cite{bbp_belle}.
No signal of the $\Uf\to[B\bar{B}]^{\pm}\pi^{\mp}$ decay is found,
with upper limit on its fraction of $<0.4\,\%$ at 90\% confidence
level.

The distributions in the $B\bar{B}^*$ and $B^*\bar{B}^*$ invariant
mass for the $\Uf\to[B\bar{B}^*]^{\pm}\pi^{\mp}$ and
$\Uf\to[B^*\bar{B}^*]^{\pm}\pi^{\mp}$ signal regions, respectively,
indicate clear excess of events over background, peaking at the
thresholds [see Fig.~\ref{fig:bbpi}~(b) and~(c)]. These threshold
peaks are interpreted as the signals of the $\zbo\to B\bar{B}^*$ and
$\zbt\to B^*\bar{B}^*$ decays, with significances of $8\,\sigma$ and
$6.8\,\sigma$, respectively. Despite much larger phase-space, no
significant signal of the $\zbt\to B\bar{B}^*$ decay is found.

Assuming that the $\zb$ decays are saturated by the channels so far
observed, Belle calculated relative branching fractions of the $\zbo$
and $\zbt$ (see Table~\ref{tab:zb_dec}).
\begin{table}[tb!h]
\caption{Branching fractions ($\br$) of $\zbo$ and $\zbt$ assuming
  that the observed so far channels saturate their decays.}
\label{tab:zb_dec}
\center
\begin{tabular}{l|c|c}
\hline
Channel & $\br$ of $\zbo$, \% & $\br$ of $\zbt$, \% \\
\hline
$\Uo\pip$  & $0.32\pm0.09$ & $0.24\pm0.07$ \\
$\Ut\pip$  & $4.38\pm1.21$ & $2.40\pm0.63$ \\
$\Uth\pip$ & $2.15\pm0.56$ & $1.64\pm0.40$ \\
$\hb\pip$  & $2.81\pm1.10$ & $7.43\pm2.70$ \\
$\hbp\pip$ & $2.15\pm0.56$ & $14.8\pm6.22$ \\
$B^+\bar{B}^{*0}+\bar{B}^0B^{*+}$ & $86.0\pm3.6$ & -- \\
$B^{*+}\bar{B}^{*0}$ & -- & $73.4\pm7.0$ \\
\hline
\end{tabular}
\end{table}
The $B^{(*)}\bar{B}^*$ channel is dominant and accounts for about 80\%
of the $\zb$ decays. The $\zbt\to B\bar{B}^*$ channel is not included
in the table because its significance is marginal. If considered, the
$\zbt\to B\bar{B}^*$ branching fraction would be
$(25.4\pm10.2)\%$. All other fractions would be reduced by a factor of
1.33.

\section{Evidence for neutral isotriplet member $\zbo^0$}

Both $\zbo$ and $\zbt$ are isotriplets with only charged components
observed originally. Belle searched for their neutral components using
the $\Uf\to\Un\pi^0\pi^0$ ($n=1,2$) decays~\cite{zb_belle_neutral}.
These decays are observed for the first time and the measured
branching fractions
$\br[\Uf\to\Uo\pi^0\pi^0]=(2.25\pm0.11\pm0.22)\times10^{-3}$ and
$\br[\Uf\to\Ut\pi^0\pi^0]=(3.66\pm0.22\pm0.48)\times10^{-3}$, are in
agreement with isospin relations.

\begin{figure}[tbhp]
\includegraphics[width=0.48\linewidth]{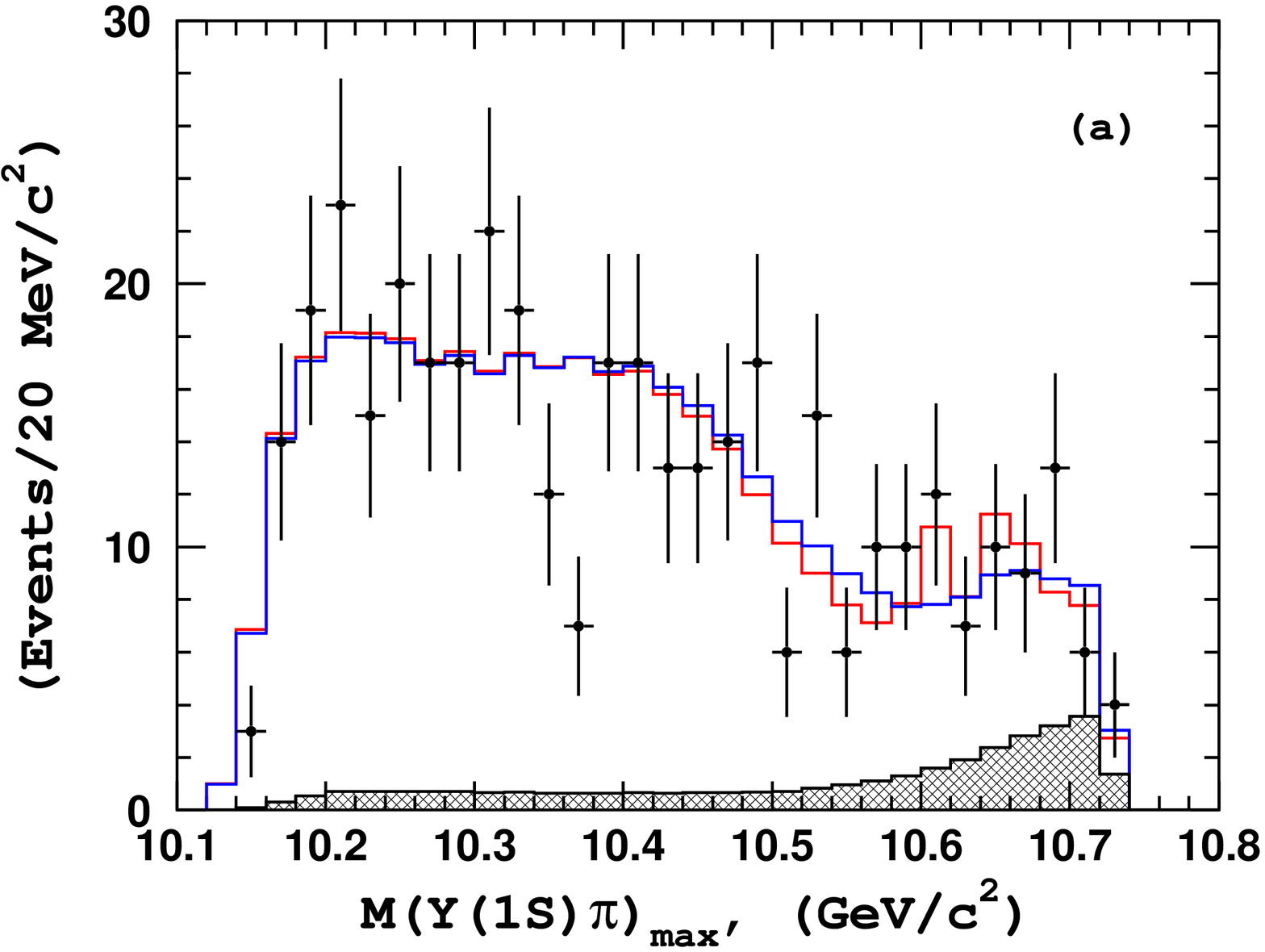}
\includegraphics[width=0.48\linewidth]{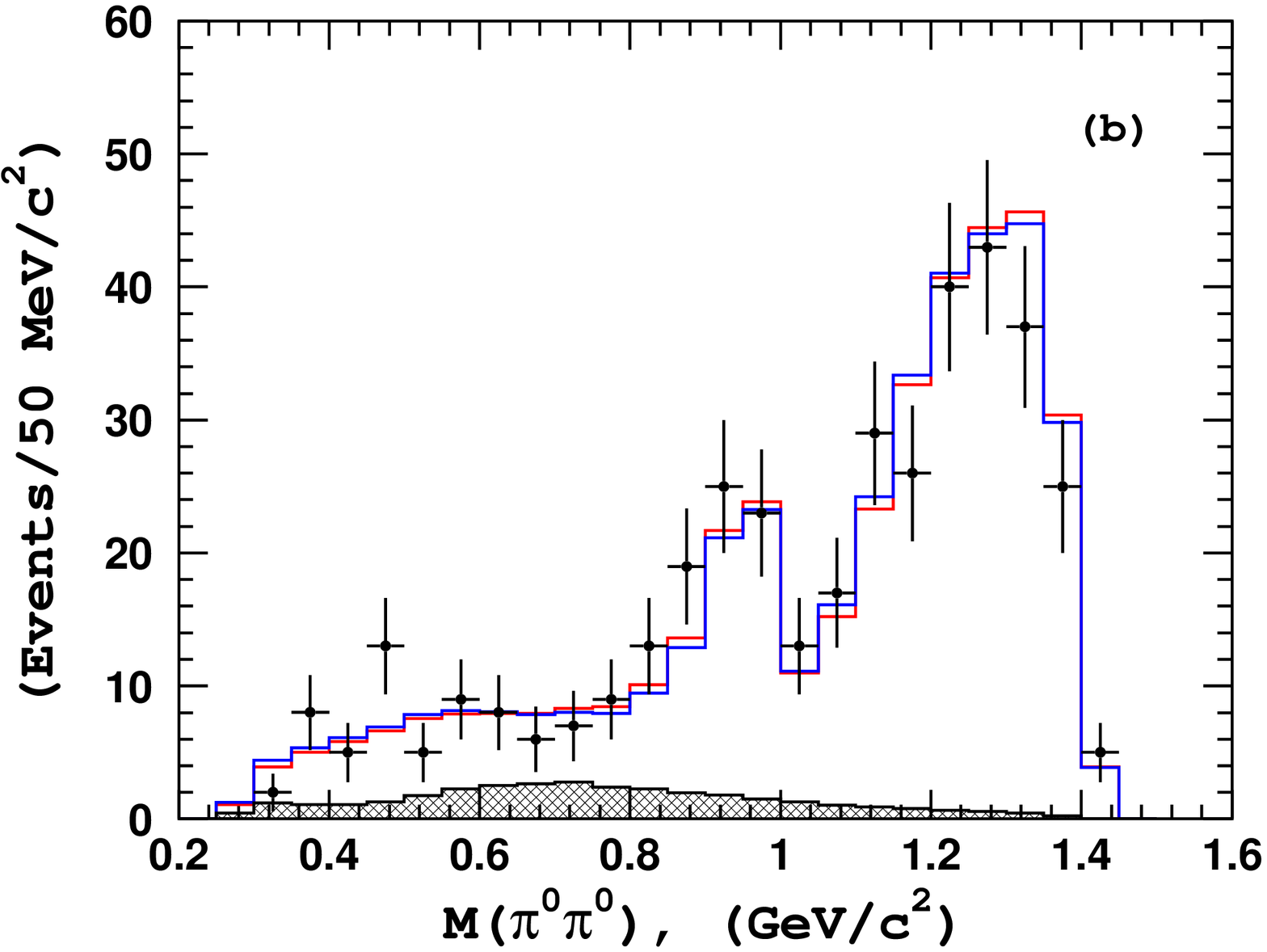}
\includegraphics[width=0.48\linewidth]{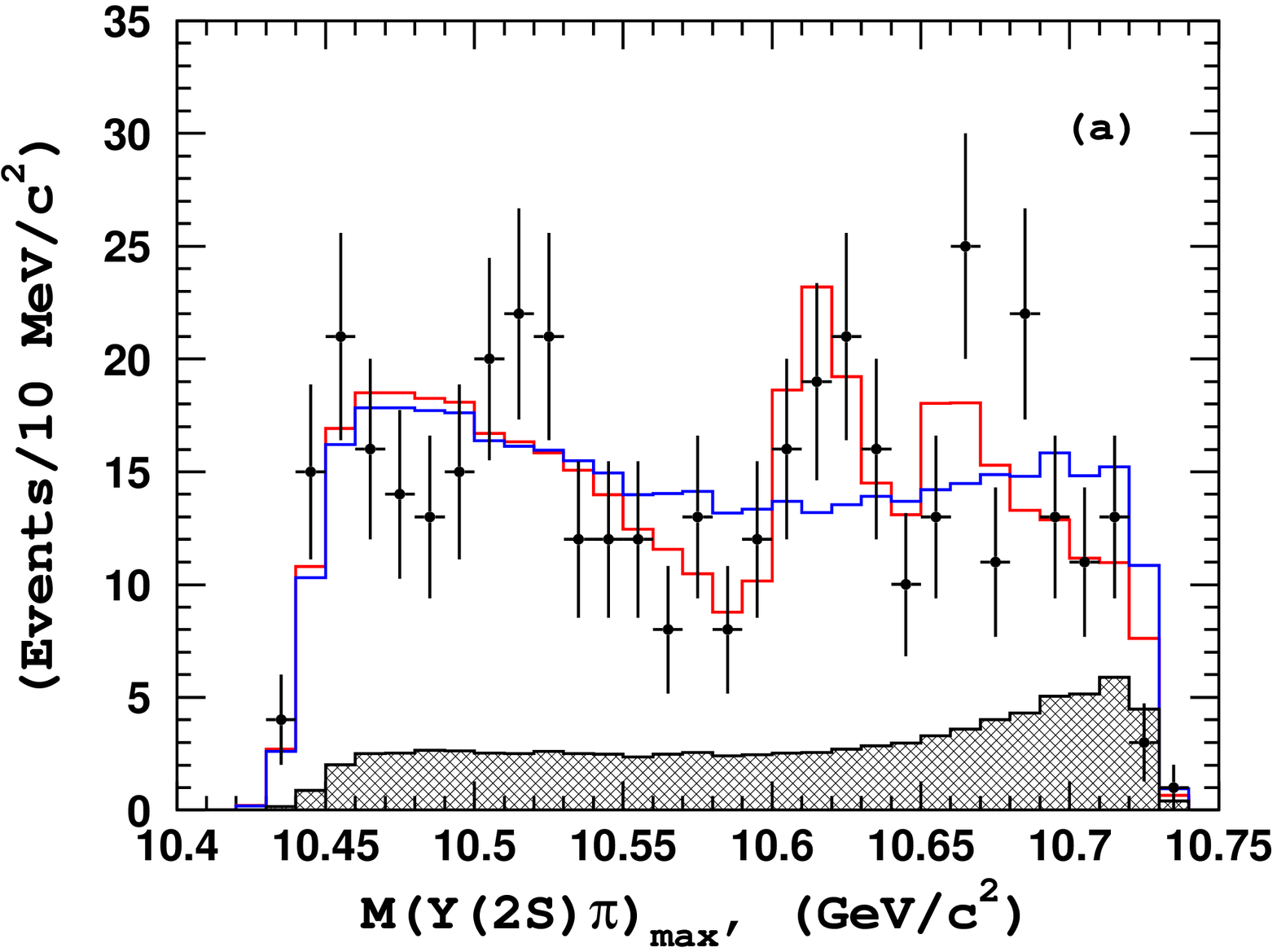}
\includegraphics[width=0.48\linewidth]{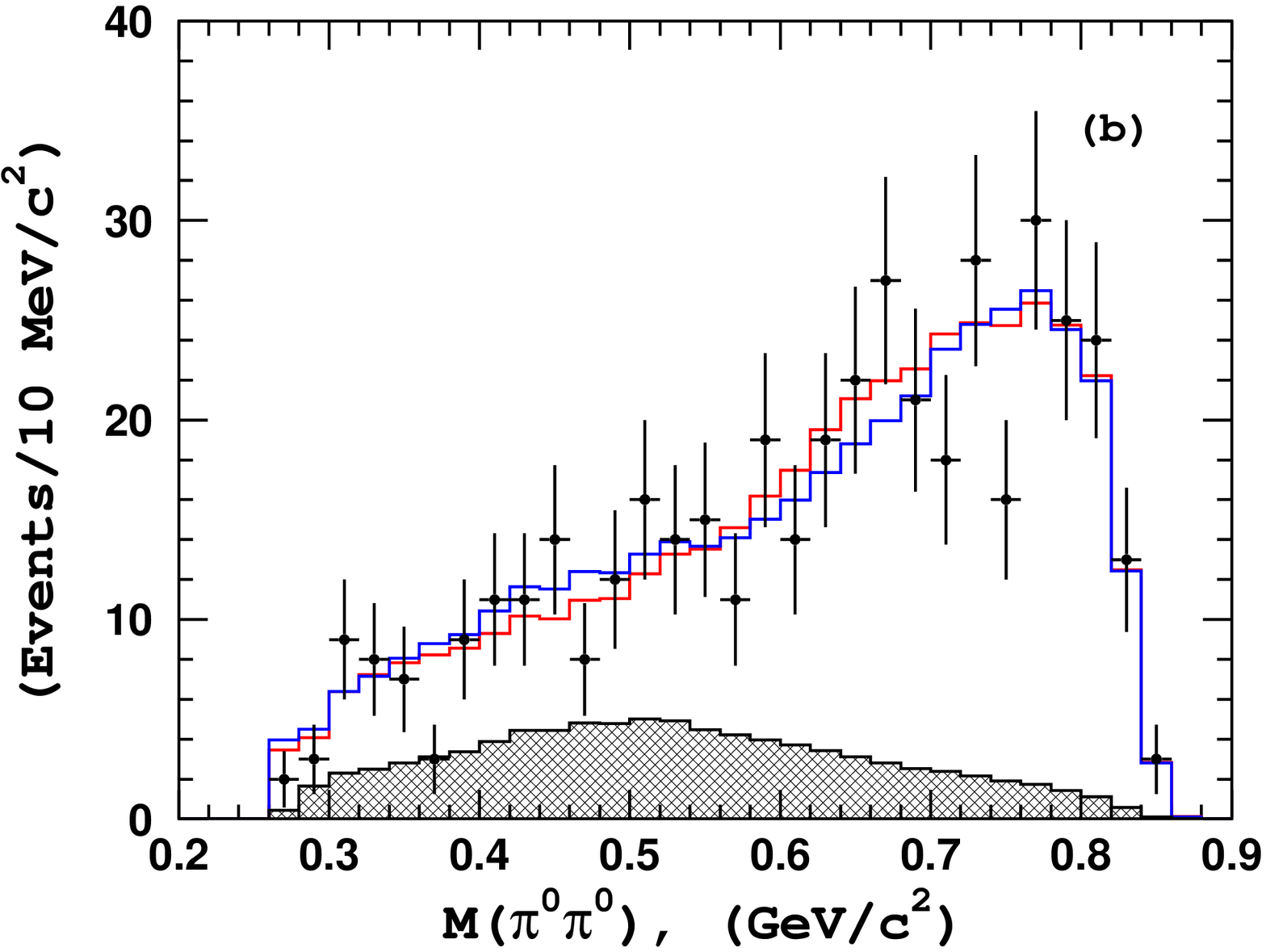}
\caption{ The projections of the Dalitz plot fit for the
  $\Uo\pi^0\pi^0p$ (top row) and $\Ut\pi^0\pi^0$ (bottom row) channels
  on the $\Un\pi^0$ (left column) and $\pi^0\pi^0$ invariant mass. }
\label{fig:zb_neu}
\end{figure}

Belle performed the Dalitz plot analyses of the
$\Uf\to\U(1S,2S)\pi^0\pi^0$ transitions using the same model as for
the charged pion channels (see Fig.~\ref{fig:zb_neu}).
The $\zbo^0$ signal is found in the $\Ut\pi^0$ channel with the
significance of $4.9\,\sigma$ including systematics. The $\zbo^0$ mass
of $(10609^{+8}_{-6}\pm6)\,\mevm$ is consistent with the charged
$\zbo^{\pm}$ mass. The signal of the $\zbo^0$ in the $\Uo\pi^0$
channel and the $\zbt^0$ signal are insignificant. The Belle data do
not contradict the existence of the $\zbo^0\to\Uo\pi^0$ and the
$\zbt^0$, but the available statistics are insufficient to establish
these signals.

\section{Interpretations}

As discussed at the end of Section~\ref{sec:obs}, the assumption of
molecular $B^{(*)}\bar{B}^*$ structure naturally explains all observed
so far properties of the $\zb$ states. Their dynamical model, however,
is an open question. Proposed interpretations include presence of the
compact tetraquark~\cite{zb_ali}, non-resonant
rescattering~\cite{zb_thresh}, multiple rescatterings that result in
the amplitude pole known as coupled channel
resonance~\cite{zb_cc_resonance} and deutron-like molecule bound by
meson exchanges~\cite{zb_molec}. All these mechanisms (except for the
tetraquark) are intimately related and correspond rather to
quantitative than to qualitative differences. Further experimental and
theoretical studies are needed to clarify the nature of the $\zb$
states.

As discussed in Ref.~\cite{zb_voloshin}, based on heavy quark symmetry
one can expect more states with similar nature but with differing
quantum numbers. Such states should be accessible in radiative and
hadronic transitions in data samples with high statistics at and above
the $\Uf$, that will be available at the SuperKEKB.

\section{Summary}

Despite observed only recently, the $\zb$ states provide a very rich
phenomenological object with a lot of experimental information
available. They could be very useful for understanding dynamics of the
hadronic systems near and above the open flavor thresholds.


\begin{thebibliography}{99}

\bibitem{x3872_belle_1st}
  S.~K.~Choi {\it et al.}  [Belle Collaboration],
  Phys.\ Rev.\ Lett.\  {\bf 91}, 262001 (2003).

\bibitem{zb_belle} 
  A.~Bondar {\it et al.}  [Belle Collaboration],
  Phys.\ Rev.\ Lett.\ {\bf 108}, 122001 (2012).

\bibitem{hb_belle}
  I.~Adachi {\it et al.} [Belle Collaboration],
  Phys.\ Rev.\ Lett.\ {\bf 108}, 032001 (2012).

\bibitem{zb_belle_angular} 
  I.~Adachi {\it et al.} [Belle Collaboration],
  arXiv:1105.4583 [hep-ex].

\bibitem{zb_voloshin} 
  A.~E.~Bondar, A.~Garmash, A.~I.~Milstein, R.~Mizuk and M.~B.~Voloshin,
  Phys.\ Rev.\ D {\bf 84}, 054010 (2011).

\bibitem{zb_belle_bb} 
  I.~Adachi {\it et al.}  [Belle Collaboration],
  arXiv:1209.6450 [hep-ex].

\bibitem{bbp_belle} 
  A.~Drutskoy {\it et al.}  [Belle Collaboration],
  Phys.\ Rev.\ D {\bf 81}, 112003 (2010).

\bibitem{zb_belle_neutral} 
  I.~Adachi {\it et al.}  [Belle Collaboration],
  arXiv:1207.4345 [hep-ex].

\bibitem{zb_ali} 
  A.~Ali, C.~Hambrock and W.~Wang,
  Phys.\ Rev.\ D {\bf 85}, 054011 (2012).

\bibitem{zb_thresh} 
  D.~-Y.~Chen and X.~Liu,
  Phys.\ Rev.\ D {\bf 84}, 094003 (2011).

\bibitem{zb_cc_resonance} 
  I.~V.~Danilkin, V.~D.~Orlovsky and Y.~.A.~Simonov,
  Phys.\ Rev.\ D {\bf 85}, 034012 (2012).

\bibitem{zb_molec} 
  S.~Ohkoda, Y.~Yamaguchi, S.~Yasui, K.~Sudoh and A.~Hosaka,
  Phys.\ Rev.\ D {\bf 86}, 014004 (2012).

\end{thebibliography}
\end{document}